# Event-Driven Market Co-Movement Dynamics in Critical Mineral Equities: An Empirical Framework Using Change Point Detection and Cross-Sectional Analysis

*Haibo Wang*

*A.R. Sánchez Jr. School of Business, Texas A&M International University, Laredo, Texas, USA*

*hwang@tamui.edu*

**Abstract**

This study examines market behavior in critical mineral investments using a novel analytical framework that combines change-point detection (PELT algorithm) with cross-sectional analysis. This research analyzes ESG-ranked critical mineral ETFs from March 31, 2014, to April 19, 2024, using the S&P 500 as a benchmark to evaluate market co-movements. The findings demonstrate that different critical mineral investments experienced change points at distinct times, but three major dates—July 23, 2015; March 17, 2020; and December 1, 2020—were common and aligned with global events such as the oil market shock, the COVID-19 pandemic, and later market adjustments. Herding behavior among investors increased after these shocks, following the 2015 and 2020 crises, but shifted to anti-herding after positive vaccine news in late 2020 and after the Russian invasion of Ukraine in 2022. The sensitivity analysis shows that investor coordination is strongest during market downturns but exhibits greater variation during stable periods or after major developments, with these dynamics sensitive to the length of the observation period. Additionally, anti-herding became more apparent during crises, suggesting investors reacted to specific risks rather than moving in lockstep, especially in response to geopolitical shocks.

JEL classifications O13; G41; C51; G14; G12

## 1. Introduction

Climate change is a pressing global challenge, driving demand for renewable energy technologies and critical minerals. Investors have seen significant returns from portfolios focused on minerals like aluminum, cobalt, copper, graphite, lithium, nickel, and rare earth elements. However, extraction of these minerals often occurs in developing nations, raising environmental, social, and governance (ESG) concerns. For a comprehensive overview of ESG scores and rankings for countries holding the largest reserves of critical minerals, please refer to Table 1.

Table 1. ESG score and rank of countries with the largest reserves of critical minerals.

| Critical Mineral | Countries with the largest reserves | Environmental Performance Index (rank)[1] | Social Progress Index (rank)[2] | Government Effectiveness Index (rank)[3] |
|---|---|---|---|---|
| Lithium | China | 28.40 (160) | 65.44 (100) | 0.84 (44) |
| Nickel | Indonesia | 28.20 (164) | 66.67 (87) | 0.38 (62) |
| Cobalt | Democratic Republic of Congo | 36.90 (119) | 42.70 (161) | -1.72 (185) |
| Graphite | China | 28.40 (160) | 65.44 (100) | 0.84 (44) |
| Copper | Zambia | 38.40 (106) | 53.29 (135) | -0.82 (154) |
| Aluminum | China | 28.40 (160) | 65.44 (100) | 0.84 (44) |
| Rare earth elements | China | 28.40 (160) | 65.44 (100) | 0.84 (44) |

Note. The Environmental Performance Index (EPI), developed by Yale and Columbia Universities, ranks countries based on environmental health and ecosystem vitality using multiple indicators. It provides comparative data to track countries' progress on sustainability and climate goals. The Social Progress Index (SPI), produced by the Social Progress Imperative, measures how well countries meet their citizens' social and environmental needs. It is a comprehensive societal performance metric independent of economic indicators. The Government Effectiveness Index (GEI), part of the World Bank's Worldwide Governance Indicators, assesses the quality of public services and the effectiveness of policy implementation across countries. It is one of six governance dimensions used for global comparisons of institutional performance.

Research suggests a positive but inconsistent correlation between ESG scores and financial performance. The relationship between ESG scores and herding behavior in financial markets is complex. Some studies show ESG scores are associated with anti-herding behavior among ESG-focused funds, potentially enhancing risk-adjusted returns (Verheyden *et al.* 2016; Daugaard 2020; Shanaev & Ghimire 2022). In the meantime, evidence also finds that herding behavior can be detected in stocks with high ESG scores, particularly in certain industries such as European banking, where herding can influence the convergence of ESG performance. Wang (2023) found significant post-2018 divergence followed by herd behavior towards distinct clusters in EU banks' ESG performance. Gavrilakis and Floros (2023) found significant post-2018 divergence followed by herd behavior towards distinct clusters in EU banks' ESG performance.

---

[1] https://epi.yale.edu/epi-results/2022/component/epi
[2] https://www.socialprogress.org/
[3] https://databank.worldbank.org/source/worldwide-governance-indicators

While the literature explores ESG and herding behavior in bank stocks, the influence of ESG scores on critical mineral investment remains unexplored, with no prior studies evaluating their impact on critical mineral portfolios. To bridge this knowledge gap, this study presents the following research questions (RQs):

RQ1. Will critical mineral portfolios with high ESG scores motivate event-driven behavior and be less sensitive to negative market shocks in the long term?

RQ2. How do extreme events and market shocks affect the short-term momentum of trading activity on critical minerals?

RQ3. What are the market co-movement dynamics among the critical mineral portfolios with different time horizons?

This study tackles research questions (RQ1-RQ3) by employing a novel empirical exploration framework that includes change point detection (CPD) using the pruned exact linear time (PELT) algorithm, and cross-sectional analysis (CSA) models using Cross-Sectional Standard Deviation (CSSD) and Cross-Sectional Absolute Deviation (CSAD) metrics for event-driven trading activities, short-term momentum, and market co-movement dynamics analysis. This framework accounts for regime shifts, asymmetric shock responses, and co-movement dynamics. The selected ETFs—LIT, REMX, COPX, PICK, VAW, and IYM—each provide targeted exposure to different critical minerals or broader metals and mining sectors, using transparent, rules-based index methodologies. The S&P 500 Index, which represents about 80% of the U.S. equity market, provides essential context for risk and return analysis by serving as a benchmark for performance and diversification benefits. This data selection encompasses daily prices and returns of critical mineral portfolios over a decade to ensure broad applicability and robust analysis. To verify the reliability of the findings, this study conducts extensive sensitivity analyses to identify potential biases and limitations within the experimental design. This includes varying the time window to assess the model's sensitivity to data irregularities and stability over time.

This study fills a knowledge gap by specifically investigating the influence of economic /geopolitical crises on critical mineral investment portfolios, areas that the literature indicates have not been previously explored. The analytical framework, which integrates PELT for CPD and cross-sectional analysis, offers advantages over traditional methods by efficiently handling large, non-stationary datasets and detecting multiple, potentially subtle structural shifts. In addition, this study evaluates the impact of different time horizons on the measurement of herding dynamics using time-series data, finding that the chosen time frame influences herding dynamics and highlighting the time-sensitive nature of these phenomena. Furthermore, this integrated approach addresses limitations of traditional single-event studies by providing a more comprehensive understanding of investor behavior, accounting for market response latency and information dynamics.

The remainder of this paper is organized into five sections. Section 2 discusses the theoretical basis of this research and provides a comprehensive review of the relevant literature. Section 3 outlines the analytical framework and methodology employed, along with a detailed description of the data sources used. Section 4 presents empirical results, while Section 5 examines the limitations of this study and explores its implications for future research and practice. Finally, Section 6 concludes with a summary of the essential findings and their significance.

## 2. Literature review and empirical background

### 2.1. Investment in critical minerals

Following the 2015 Paris Agreement, investments in renewable and clean energy technologies have increased by 10 percentage points, from 2% to 12%. Particularly noteworthy is that over 80% of total investments in the power sector have been allocated to renewable, grid, and storage technologies, such as solar PV, batteries, and electric vehicles, all of which rely on critical minerals as essential components[4]. The literature on critical mineral investment is extensive. Research has examined price persistence in critical minerals for new-energy vehicle production (Claudio-Quiroga *et al.* 2023). The critical minerals industry is under growing pressure to ensure responsible exploration and production practices, as concerns about safety and environmental impact within the supply chain mount. To mitigate these risks, companies must adopt robust Environmental, Social, and Governance (ESG) standards and prioritize worker safety and community well-being. While research has established a positive link between ESG performance and financial success (Dmuchowski *et al.* 2023), it has also shown that ESG disclosure can drive technological innovation (Chen *et al.* 2023). The impact of ESG scores on risk management and long-term critical minerals investment strategies remains to be determined. Furthermore, there is a significant

---

[4] https://www.iea.org/reports/world-energy-investment-2022/overview-and-key-findings, Accessed August 30, 2024

knowledge gap regarding how ESG considerations affect the extraction and production of critical minerals and their associated supply chains, underscoring the need for further research and industry innovation.

*2.2. Impact of the crisis on investor behavior*

Several studies have examined herding behavior in various markets during times of crisis. During the COVID-19 pandemic, Raggad (2023) found that investors tended to follow the crowd (herding behavior) in energy commodities and ethical energy investments, with West Texas Intermediate (WTI) oil prices significantly influencing this trend rather than just correlating with it. While simpler analytical methods might not detect this crowd-following in individual commodity sectors, more advanced approaches that account for changes over time reveal it in almost all sectors, especially during and after the global financial crisis (GFC). Gabbori *et al.* (2021) observed significant herding in the Saudi stock market, noting that this behavior persisted after the GFC, even after accounting for market movements that might *appear* to be related by chance. Similarly, Rubbaniy *et al.* (2025) found significant herding in North American energy stocks that persisted through periods such as the GFC and COVID-19. They also identified key areas for further research, including a need for more studies on how external events dynamically affect herding in commodity markets using sophisticated analytical methods. Specifically, they highlighted the importance of distinguishing between herding driven by actual economic value and that driven by speculation in energy stock markets, and the current limitations in fully capturing herding caused by investor sentiment or emotion.

While these studies suggest that herding behavior is a common phenomenon in various markets during times of crisis, others have found empirical evidence of anti-herding behavior during crisis (Babalos & Stavroyiannis 2015; Stavroyiannis & Babalos 2020). While the literature investigates the impact of the crisis on herding and anti-herding behavior, primarily in the oil market and energy commodities, the influence of the crisis on critical mineral investment remains underexplored, with no prior studies evaluating its impact on critical mineral portfolios, which are key components of renewable energy development.

*2.3. Change Point Detection Methods*

Change point detection (CPD) is crucial for identifying structural shifts in financial time series, such as market shocks and regime changes (Aminikhanghahi & Cook 2017). Traditional methods like CUSUM and likelihood ratio tests are limited: CUSUM assumes constant variance (Fearnhead & Rigaill 2019) and can yield false signals in volatile markets (Wang & Ning 2025). Likelihood ratio tests require strong distributional assumptions (Kim & Siegmund 1989), struggle with multiple or gradual changes (Kim 1996), and are computationally intensive for long datasets (Wang & Xie 2024). Bayesian approaches provide probabilistic modeling and can incorporate prior knowledge (Barry & Hartigan 1993), but they are computationally demanding (Ruggieri & Antonellis 2016) and less effective at capturing gradual shifts (Wilson *et al.* 2010) or handling high-dimensional data (Xie *et al.* 2013). Binary segmentation, while simple, can miss closely spaced (Piotr 2014) or subtle changes (Matteson & James 2014) and lacks global optimality (Fryzlewicz 2020).

In comparison, the Pruned Exact Linear Time (PELT) algorithm offers several advantages (Maidstone *et al.* 2017). This study is especially relevant for long-term financial datasets, such as the cumulative returns of critical mineral ETFs over a decade, which show varying degrees of interconnectedness based on mutual information results. PELT efficiently detects multiple change points by minimizing a global cost function (Killick *et al.* 2012) and adapts well to large samples, ensuring precise segmentation even when structural similarities or differences exist among the

ETFs (Haynes *et al.* 2017). Its ability to handle complex, non-stationary time series without stringent distributional or independence assumptions makes it well-suited for analyzing the long-term co-movement and regime shifts in diverse critical mineral portfolios (Killick *et al.* 2012). Researchers report the studies on the co-movement of gasoline and diesel prices. Magazzino *et al.* (2024) used neural networks to analyze EU fuel prices, suggesting a latent factor drives their co-movement. This unobserved component appears to influence short-run price forecasting. This analysis of European fuel prices suggests a strong long-term co-movement between gasoline and diesel. Surprisingly, this relationship appears unaffected by differing national fuel tax systems (Mutascu *et al.* 2022).

## 3. Methodology and model

*3.1. Data and variables*

This study examines exchange-traded funds (ETFs) representing critical minerals with the highest ESG scores as primary assets from March 31, 2014, to April 19, 2024, using the earliest date for complete data across all seven ETFs. Price data for these ETFs, covering minerals such as lithium, rare earths, nickel, cobalt, graphite, aluminum, and copper, were sourced from the Center for Research in Security Prices (CRSP) database. The S&P 500 Index, sourced from S&P Global, is the broad market benchmark to assess co-movement and event-driven effects on critical mineral equities. ETFs were selected over individual stocks for their diversification benefits, standardized data, and ability to capture segments across the global minerals value chain, including companies not listed in the U.S.

Figures 1 and 2 illustrate the performance trajectory of seven critical mineral portfolios with high ESG scores from March 31, 2014, to April 19, 2024. The overall pattern shows an upward trend, with noticeable drawdowns. A significant drawdown on February 20, 2020, suggests a potential correlation with the onset of the COVID-19 pandemic. Another minor drawdown deviation on September 28, 2022, raises the possibility of a connection to the Nord Stream sabotage incident. Additionally, a minor drawdown variation aligns with the 2014-2016 global economic slowdown, characterized by a significant decline in oil prices. Insights from Agoraki *et al.* (2023) highlight the effects of the COVID-19 outbreak on market sentiment and green investment funds, offering valuable context for interpreting the observed performance dynamics in Figures 1 and 2. These rapid changes indicate the possibility of investor trading activities and short-term momentum.

Figure 1. The daily adjusted closing price of seven critical mineral equities from March 31, 2014, to April 19, 2024.

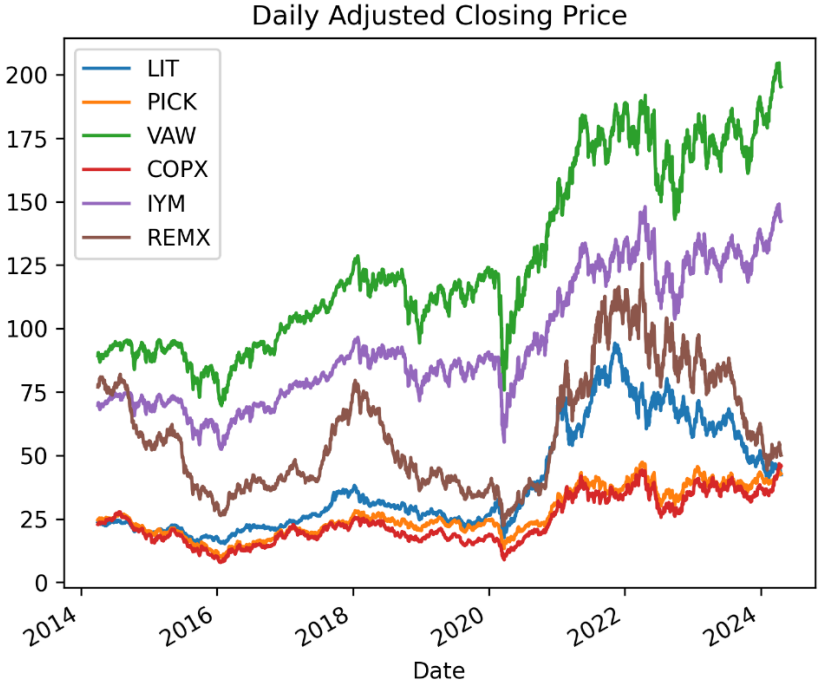

Figure 2. The cumulative returns of seven critical mineral equities from March 31, 2014, to April 19, 2024.

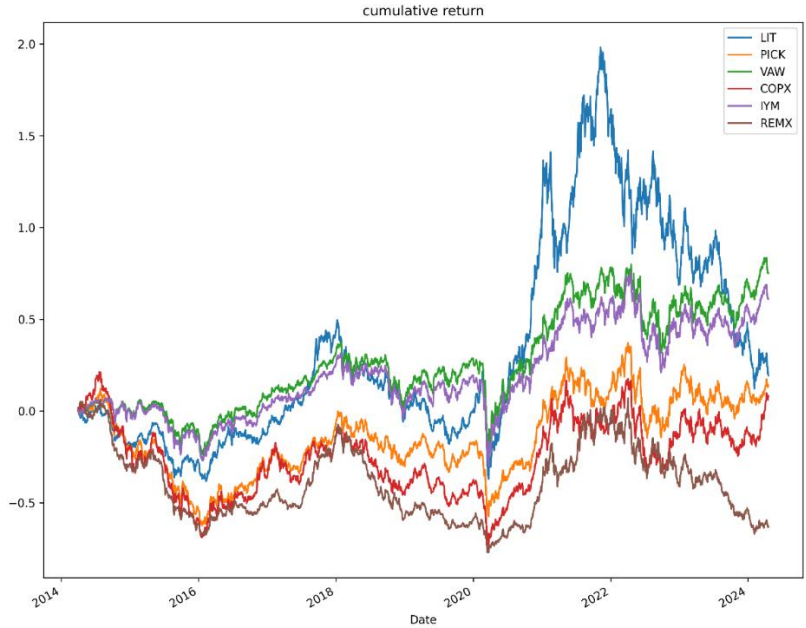

The variable descriptions are given in Table 2. Table 3 reports the ESG scores for seven portfolios focused on critical mineral investments, sourced from MSCI. IYM (aluminum) and VAW (cobalt) have higher ESG scores than other portfolios.

This study splits the data into multiple samples according to the crisis timeline. The first event occurred on February 22, 2022, when Russia invaded Ukraine, exposing the vulnerabilities of countries that heavily rely on Russian oil and gas. This event had a negative impact on the global energy market. The second event is the start of the Middle East crisis on October 7, 2023. This event had a negative impact on the global energy market. These samples allow for a focused examination of the event-driven dynamics.

Table 2. Descriptions of the variables

| Variable | Symbol | Description | Source |
| --- | --- | --- | --- |
| Portfolios on critical minerals | LIT, PICK, VAW, COPX, IYM, REMX | ETFs for lithium, nickel, cobalt, copper, aluminum, rare earth elements, and key components of renewable energy equipment | CRSP |

| | | | | |
|---|---|---|---|---|
| S&P500 Index | SPX | The index comprises 500 prominent companies, representing roughly 80% of the available market capitalization.[5] | | S&P |

Note. These portfolios capture the performance and characteristics of companies across various critical minerals value chain segments. LIT provides targeted exposure to the full lithium cycle, from mining and refining the metal to battery production. REMX focuses on companies that produce, refine, and recycle rare earth and strategic metals and minerals. COPX offers targeted exposure to copper mining companies. Depending on its holdings and index methodology, PICK provides broader exposure to global companies involved in the extraction and production of base metals, such as nickel, and mining. VAW & IYM are broader materials sector ETFs that include major producers of industrial metals like aluminum and cobalt, as well as potential companies involved in their supply chains.

Table 3. ESG score of seven portfolios for critical minerals

| Name of mineral | Portfolio | ESG score[6] | ESG Score Peer Percentile (%) | ESG Score Global Percentile (%) |
|---|---|---|---|---|
| Lithium | LIT | 5.74 | 36.84 | 35.44 |
| Nickel | PICK | 6.0 | 44.74 | 40.15 |
| Cobalt | VAW | 6.48 | 60.53 | 51.9 |
| Copper | COPX | 6.04 | 46.49 | 40.95 |
| Aluminum | IYM | 6.55 | 64.04 | 54.59 |
| Rare earth elements | REMX | 5.07 | 25.83 | 28.95 |

Note. The selected ETFs exhibit moderate to low-moderate ESG scores, with most performing below the median compared to their specific peer groups and the broader global universe of funds. VAW and IYM stand out with relatively better ESG performance, exceeding their peer and global medians. On the other hand, REMX shows the weakest ESG profiles in this selection, and China is the key source country of rare earth elements and graphite. This pattern might reflect the inherent ESG challenges in the mining and materials sectors of countries with the most reserves these ETFs track.

*3.2. Theoretical foundations and empirical exploration framework*

3.2.1 Linking ESG scores and herding behavior to market co-movement dynamics

---

[5] https://www.spglobal.com/spdji/en/indices/equity/sp-500/#overview
[6] https://www.msci.com/indexes/ishares

Herding behavior, where investors mimic the trades of others rather than relying on their own analysis, can significantly impact asset prices and market volatility. In ESG investing, herding may be influenced by the growing emphasis on sustainable and responsible investment practices. Studies have shown that herding behavior is present in ESG markets, with investors collectively moving towards or away from ESG-rated assets based on prevailing sentiments and information (Gavrilakis & Floros 2023). Research indicates significant herding behavior in U.S. ESG leader stocks, especially during periods of market stress such as the Global Financial Crisis and the COVID-19 pandemic (Rubbaniy *et al.* 2021).

Event-driven trading involves strategies that exploit pricing inefficiencies driven by corporate events, economic releases, or geopolitical developments. Incorporating ESG-related events into this framework requires understanding how such events influence investor behavior and market dynamics (Ben Ameur *et al.* 2024). For example, positive ESG news may attract increased investor attention, leading to heightened trading activity and potential herding, which can, in turn, affect stock price volatility. Conversely, negative ESG events might trigger collective selloffs, amplifying market fluctuations.

ESG scores and related events influence investor behavior, leading to herding phenomena that impact trading patterns and market volatility. ESG disclosures and news are information catalysts, shaping investor perceptions and actions (Friede 2019). Investors influenced by ESG information may exhibit herding behavior, collectively adjusting their portfolios in response to ESG signals. The aggregated trading behavior resulting from herding can lead to increased volatility, as markets rapidly reprice assets based on the prevailing ESG sentiment (Zhang 2022).

To address research questions (RQ1-RQ3) on ESG investing and market behavior, it is essential to develop effective analytical models that account for the behavioral dynamics driven by ESG considerations.

3.3.2 The research design

The empirical exploration framework in this study consists of three components (see Figure 3): a change-point detection analysis to identify significant changes in relationships or dependencies within time-series data, a cross-sectional analysis of herding behavior in the critical mineral market, and a sensitivity analysis across various time windows.

This study applies PELT to analyze returns of critical mineral portfolios over a decade. The algorithm's linear computational complexity and ability to process millions of data points efficiently make it ideal for detecting structural breaks, market manipulation, and fraud in this large-scale financial time series. PELT's superiority in handling abrupt and subtle changes and its scalability and precision position it as a powerful tool for enhancing regulatory oversight and ensuring market integrity (Bai & Perron 2003; Killick *et al.* 2012). The detailed formulation and mathematical proof of PELT are provided in the Appendix.

Figure 3. Analytics Framework Using CPD and CSA

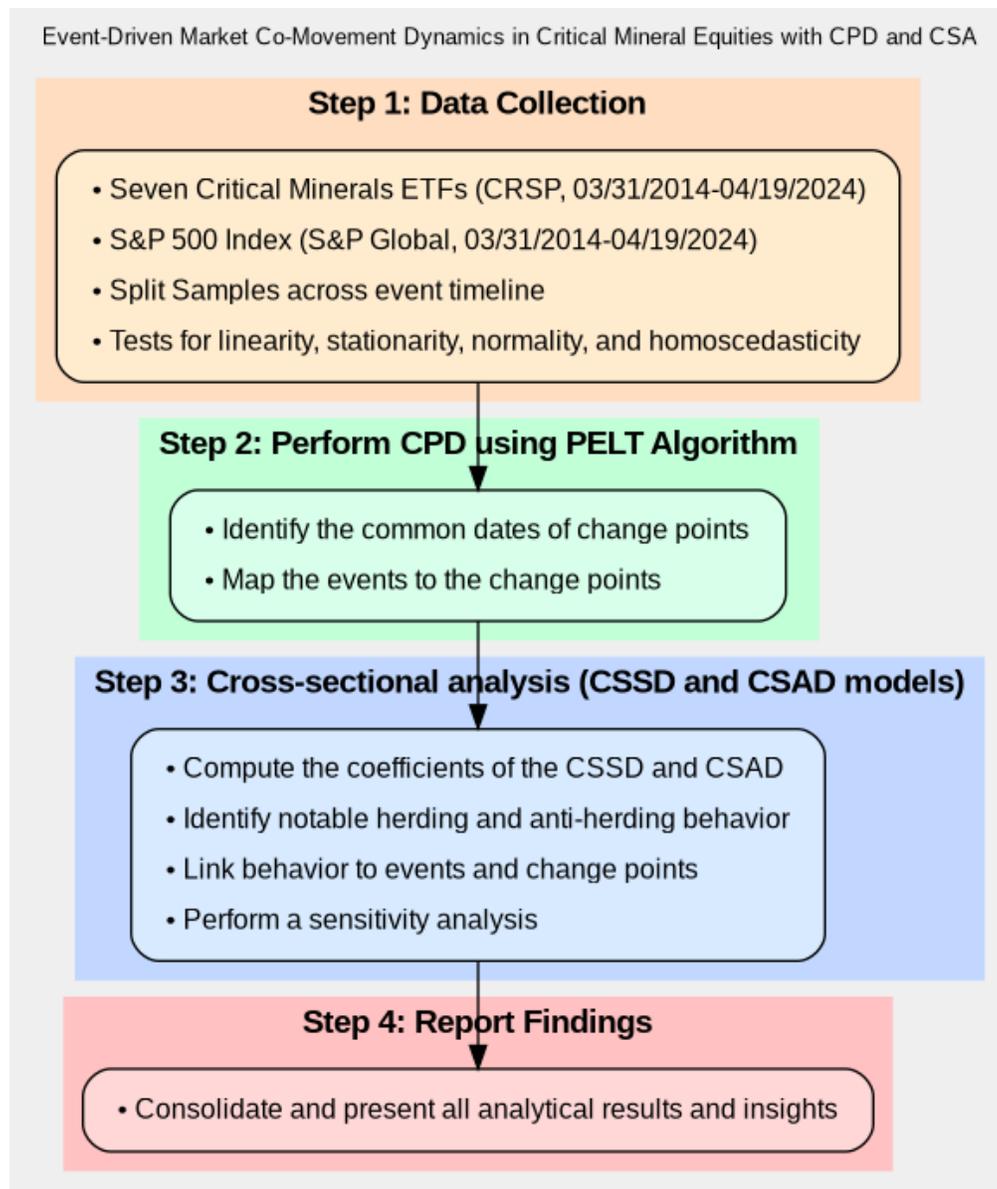

To examine herding behavior in financial markets, particularly in the context of critical mineral equities, this study employs two key metrics: CSSD and CSAD. These metrics provide valuable insights into the degree of dispersion and absolute deviation in equity returns, allowing for a more nuanced understanding of herding dynamics in critical mineral markets (Christie & Huang 1995; Chang & Lin 2015). The CSSD metric measures the dispersion of individual behavior or opinions within a group, while the CSAD metric estimates the firm-level volatility of individual assets within a cross-sectional framework. The CSSD and CSAD metrics are used to investigate the influence of herd behavior during periods of significant market movements, and the results can provide empirical evidence supporting the Information Cascades Theory and the Reputational Herding Theory (Scharfstein & Stein 1990; Avery & Zemsky 1998; Bikhchandani *et al.* 1998). The study introduces a modified CSAD metric that accounts for both upward and downward market conditions, and the results can indicate the presence of herding and changes in investor behavior over time. The core assumptions of regression models, CSSD, and CSAD require the use of stationary data for cumulative returns to ensure the validity and reliability of the outcomes. Detailed formulations of the CSSD and CSAD are provided in the Appendix.

By combining PELT and Cross-sectional Analysis (CSA), this research develops a powerful approach for analyzing time-series data to understand trading behavior and identify the latency of the market's response to extreme events. This integrated approach can provide valuable insights into the underlying mechanisms driving market dynamics and inform decision-making in finance and economics. All methods are implemented in the Python *statsmodels* library.

## 4. Empirical results

The CPD and CSA models for time series data are based on several underlying assumptions, including linearity, stationarity, normality, and homoscedasticity among independent variables. Therefore, before analyzing the data using the proposed framework, this research examines the data's statistical properties and applies techniques to transform the variables if the assumptions are violated. The analysis begins with a report of the descriptive results for the critical mineral equities, followed by a CPD analysis, and then the application of the CSA models.

*4.1 Descriptive results*

The statistical summary in Table 4a shows that the mean, median, and variance of the cumulative returns of the critical mineral equities are approximately zero for the entire period. The small negative skewness values suggest a slightly asymmetrical and left-skewed distribution. The kurtosis values for PICK, IYM, and REMX indicate that the distributions are normal. At the same time, LIT exhibits slightly more peaked distributions, while COPX displays a slightly flatter distribution. Overall, the cumulative returns largely satisfy the normal distribution requirement. This study implements the ARCH test for homoscedasticity and the Jarque-Bera test for normality on the critical mineral equities. Table 4b reports the LM and Jarque-Bera statistics. The critical mineral market displays two key statistical characteristics. First, these minerals exhibit heteroskedasticity, meaning price volatility varies significantly across periods. Second, price distributions show fat tails, meaning extreme price movements occur more frequently than expected in a normal distribution.

Table 4a. Statistical Summary of critical mineral equities' cumulative returns

| Ticker | Mean | Median | Variance | Kurtosis | Skewness |
| --- | --- | --- | --- | --- | --- |
| LIT | 7.40E-05 | 7.21E-04 | 7.87E-04 | 5.33*** | -0.18*** |
| PICK | 5.35E-05 | 2.60E-04 | 2.26E-04 | 2.31*** | -0.35*** |
| VAW | 2.94E-04 | 9.32E-04 | 2.65E-04 | 4.28*** | -0.49*** |
| COPX | 3.24E-05 | 4.06E-04 | 2.40E-04 | 1.52*** | -0.24*** |
| IYM | 2.39E-04 | 7.20E-04 | 2.50E-04 | 3.59*** | -0.47*** |
| REMX | -2.55E-04 | -3.76E-05 | 1.75E-04 | 2.94*** | -0.21*** |

Note: (. p-value<=0.1, * p-value<=0.05, ** p-value<=0.01, ***, p-value<=0.005). The return distributions of critical mineral ETFs differ from a normal distribution, with significant negative skewness indicating a higher likelihood of large losses than gains. This increases downside risk and can limit long-term growth. Some ETFs (LIT, VAW, IYM) show high kurtosis, meaning their returns are more volatile with frequent extreme changes, resulting in less predictable cumulative returns. Others (PICK, COPX) have lower kurtosis, leading to smoother cumulative performance. REMX, in particular, has an average negative return and negative skewness, resulting in a consistent downward trend in cumulative returns.

Table 4b. ARCH test for homoscedasticity and Jarque-Bera test for normality of critical mineral equities' cumulative returns

| Ticker | ARCH Test LM Statistic | Jarque-Bera Statistic |
|---|---|---|
| LIT | 501.5396*** | 3700.7120*** |
| PICK | 589.8495*** | 3284.8824*** |
| VAW | 824.5919*** | 9580.0158*** |
| COPX | 636.8027*** | 2625.7089*** |
| IYM | 679.5354*** | 6158.8101*** |
| REMX | 461.0196*** | 1551.5617*** |

Note: (. p-value<=0.1, * p-value<=0.05, ** p-value<=0.01, ***, p-value<=0.005). The LM and Jarque-Bera statistics show two key things about the critical mineral equities. First, their volatility, or how much their prices swing, isn't steady; it changes over time, with periods of wilder swings and smooth movements. Second, their price movements don't follow the typical smooth "bell curve" pattern; big price jumps, both up and down, happen more often than such a pattern would predict.

To assess the stationarity of both average and individual asset returns, an Augmented Dickey-Fuller (ADF) test was conducted. The results in Table 5 report negative ADF t-test values and p-values of less than 0.005, providing strong evidence to reject the null hypothesis of a unit root. This suggests that the time series data is likely to be stationary, a desirable outcome as it enables the application of standard statistical techniques for analysis. The absence of a unit root indicates that the data is suitable for modeling within the proposed framework, allowing for reliable inferences to be drawn.

Table 5. Stationarity test for critical mineral equities' cumulative returns

| Ticker | ADF test (Lag) | p-value for ADF |
|---|---|---|
| LIT | -13.87***(10) | 6.49E-26 |
| PICK | -16.21***(8) | 3.97E-29 |
| VAW | -16.36***(8) | 2.89E-29 |
| COPX | -45.57***(0) | 0 |
| IYM | -16.55***(8) | 1.93E-29 |
| REMX | -13.21***(14) | 1.04E-24 |

Note: (. p-value<=0.1, * p-value<=0.05, ** p-value<=0.01, ***, p-value<=0.005). The ADF test results confirm that all critical mineral ETF cumulative return series maintain stable statistical properties over time. All seven ETFs show test statistics that are significantly negative with extremely small p-values (all below 0.0001), thus confidently reject the idea that these financial time series are nonstationary. The test required different numbers of past values (lags) for each ETF, ranging from 0 to 14, to account for their unique patterns. This consistent stability across all ETFs means their average behavior doesn't change over time, making them suitable for standard time series modeling approaches in this study without needing additional mathematical adjustments.

This study uses the mutual information function to assess the linearity between pairs of time series. Detailed formulations of the mutual information function are provided in the Appendix. By measuring the strength of the relationships between variables, mutual information provides a quantitative assessment of their dependence. Table 6 shows that the long-term performance of these critical mineral investments isn't random; they have different degrees of interconnectedness. Investments in broader materials or traditional mining sectors tend to move in the same direction because they are affected by similar economic trends. More specialized investments, like those in lithium or rare earths, show some connection to each other and the broader sector, but also have unique performance drivers. Understanding these relationships is important for building a diversified investment portfolio, as it helps to see which investments might rise or fall together versus those that behave more independently over time.

Table 6. Mutual information test for critical mineral equities' cumulative returns

| | COPX | IYM | LIT | PICK | REMX | VAW |
|---|---|---|---|---|---|---|
| **COPX** | | 0.36 | 0.21 | 0.84 | 0.33 | 0.34 |
| **IYM** | | | 0.32 | 0.47 | 0.23 | 1.82 |

|      | LIT | PICK | VAW | COPX | IYM | REMX |
|------|-----|------|-----|------|-----|------|
| **LIT**  |     |      |     | 0.27 | 0.46 | 0.30 |
| **PICK** |     |      |     |      | 0.32 | 0.47 |
| **REMX** |     |      |     |      |      | 0.23 |
| **VAW**  |     |      |     |      |      |      |

Note. The results show how much the' long-term performance (cumulative returns) of these critical mineral investments is related. Some investments show strong connections in their performance paths, like IYM and VAW, which had the highest shared information (1.82), meaning their overall gains or losses tended to move very closely together. Similarly, COPX and PICK also showed a strong link (0.84). Other pairs had moderate connections; for instance, LIT and REMX shared a reasonable amount of information (0.46), suggesting their long-term trends are somewhat similar. Many other pairs, like COPX and LIT (0.21), showed weaker links, meaning their overall performance paths were more independent.

Conditional correlation using Kendall's tau indicates that daily performance ranks of critical mineral ETFs tend to move together, even across different types of minerals. This approach shows that positive relationships are strongest among broadly focused ETFs such as REMX for multiple rare earth elements, but also exist among specialized ones, suggesting shared market drivers. Table 7 confirms all positive correlations, with the highest (0.892) between IYM and VAW, consistent with earlier mutual information findings in Table 6.

Table 7. Conditional correlation results using Kendall's tau

|      | LIT | PICK | VAW | COPX | IYM | REMX |
|------|-----|------|-----|------|-----|------|
| **LIT**  |     | 0.600*** | 0.597*** | 0.587*** | 0.601*** | 0.780*** |
| **PICK** |     |          | 0.766*** | 0.894*** | 0.791*** | 0.677*** |
| **VAW**  |     |          |          | 0.689*** | 0.984*** | 0.586*** |
| **COPX** |     |          |          |          | 0.720*** | 0.675*** |
| **IYM**  |     |          |          |          |          | 0.596*** |
| **REMX** |     |          |          |          |          |          |

Note: (. p-value<=0.1, * p-value<=0.05, ** p-value<=0.01, ***, p-value<=0.005). The results show how consistently these critical mineral ETFs' daily performance (returns) move in the same direction, rank-wise. All the numbers are positive and highly significant, meaning that when one ETF has a relatively good day (higher ranked return), the others also tend to have a good day, and similarly for bad days. The strength of this tendency varies: for example, VAW and

IYM have a very high value (0.984), indicating their daily performance rankings are almost always in the same direction. Other pairs, like PICK and COPX, also show strong agreement (0.894). Even pairs with lower values, like VAW and REMX (0.586), still show a clear, statistically reliable pattern of performance rankings moving together, just not as tightly as the higher-scoring pairs.

4.2. *The results of change point detection*

The results of change-point detection in Table 8 reveal that each equity has distinct change-point dates, with three common dates emerging: 2015-07-23 (6), 2020-03-17 (7), and 2020-12-01 (7). The possible associated event for 2015-07-23 was the shock from the price drop in the world oil market due to high supplies and a crash in the Chinese stock market (July 8-9, 2015). The event associated with 2020-03-17 was the COVID-19 pandemic, declared by the World Health Organization (WHO) on March 11, 2020. The timeline for 2020-12-01 might be related to market adjustment.

The CPD results report the latency of individual equities' responses to shocks. This latency may be attributed to short-term momentum in trading activity driven by herding or anti-herding behaviors, which motivates us to investigate this phenomenon using cross-sectional analysis.

Depending on the PELT parameters used to partition the time series into segments, the number of change points per equity can vary. It is impractical to investigate all change points on all equities. Therefore, these three common dates, along with the event dates of two geopolitical crises on 2022-02-22 and 2023-10-07, are used to examine trading activity for herding dynamics.

Table 8. Change point detection timeline of the PELT algorithm

| Ticker | CPD Timeline |
|---|---|
| LIT | 2015-07-23, 2020-03-17, 2020-12-01 |
| PICK | 2015-07-23, 2020-03-17, 2020-12-01 |
| VAW | 2015-07-23, 2020-03-17, 2020-12-01 |
| COPX | 2015-07-23, 2020-03-17, 2020-12-01 |

| IYM | 2015-07-23, 2020-03-17, 2020-12-01 |
| REMX | 2015-07-23, 2020-03-17, 2020-12-01 |

Note. The significant price shifts for most critical mineral investments in July 2015 line up well with a significant drop in Chinese stocks and global oil prices, pointing to worries about the world economy and weaker demand for raw materials. The worldwide financial panic caused a significant, across-the-board price shift in March 2020 as the COVID-19 pandemic hit. Another widespread price change in December 2020 likely happened because of good news about COVID-19 vaccines, hopes for economic recovery, and growing excitement about the need for these minerals in green energy technologies.

4.3 *CSA based on CPD and extreme event timelines*

For the entire period under consideration, the critical mineral equities show the pattern of herding behaviors using the S&P 500 index as the aggregate market return. Table 9 reports the CSSD and CSAD metrics comparing the returns of seven critical mineral equities and the S&P 500. The graphical results of the CSSD and CSAD metrics are presented in Figure 4.

Table 9. CSR for Critical mineral equities over the entire period under consideration (March 31, 2014, to April 19, 2024)

| **CSSD** | | | |
|---|---|---|---|
| | | $\beta^L$ | $\beta^U$ |
| **S&P500** | Coefficient | -0.504 | 1.159 |
| | t-stat | -19.175*** | 44.553*** |
| **CSAD** | | | |
| | | $\gamma_3$ | |
| **S&P500** | Coefficient | -0.652 | |
| | t-stat | -36.087*** | |

Note: (. p-value<=0.1, * p-value<=0.05, ** p-value<=0.01, ***, p-value<=0.005). The results indicate that investors in critical mineral stocks tend to "follow the crowd" (herd) throughout the period, especially during significant overall market movements, as shown by the CSAD results, which indicate that these stocks cluster more closely. Furthermore, these stocks behave differently in up-versus-down markets: they bunch together more when the market falls but spread out more when they rise, as the CSSD findings revealed.

Figure 4. The trend of CSSD and CSAD metrics compared to the S&P 500 from March 31, 2014, to April 19, 2024.

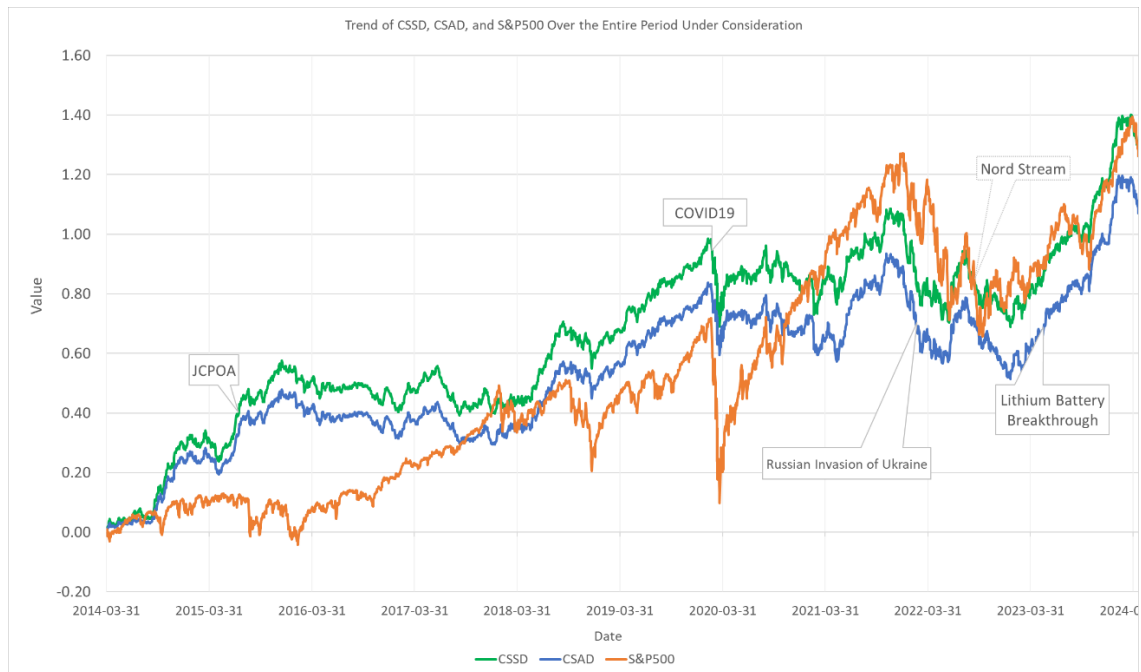

The CSSD metric shows statistically significant negative values of $\beta^L$, with a t-test value of -17.685 for the S&P500, indicative of herding behavior and irrational investor decisions during the entire period of renewable energy development. Lower $CSAD_t$ $\gamma_3$ The value indicates greater consensus or herding among critical mineral equities, suggesting that investors are closely aligned in their actions. The CSAD metric exhibits a significant negative value of $\gamma_3$, with a t-test value of 34.481 for the S&P 500, pointing to herding behavior and irrational investor choices throughout the period. These results indicate that investors in critical mineral stocks tend to herd throughout the period, especially during significant overall market movements, as shown by the CSAD results, where these stocks cluster more closely. Furthermore, these stocks behave differently in up-versus-down markets: they cluster together more when the market falls but spread out more when the market rises, as suggested by the CSSD findings.

This study divided the data into three samples based on the CPD timeline to examine investor trading behavior and short-term momentum. Table 10 reports the CSSD and CSAD metrics before and after each CPD timeline. Before the market trouble in mid-2015, investors in critical mineral stocks didn't show strong signs of herding, but this behavior became very noticeable afterwards. This herding had already occurred before the COVID-19 pandemic hit in March 2020, and it might be linked to Indonesia's Nickel Ore Export Ban in January 2020, which continued during the initial crisis. However, a significant change happened around December 2020, when good news about vaccines emerged. Instead of herding, investors began doing the opposite (anti-herding), meaning their stocks spread out more during big market moves.

Table 10. Results of CSA based on CPD timeline (07/23/2015, 3/17/2020, and 12/01/2020)

| CSSD | | Before 07/23/2015 | | After 07/23/2015 | |
|---|---|---|---|---|---|
| | | $\beta^L$ | $\beta^U$ | $\beta^L$ | $\beta^U$ |
| S&P500 | Coefficient | -0.228 | 0.277 | -0.628 | 1.343 |
| | t-stat | -39.511*** | 78.099*** | -25.931*** | 56.306*** |
| | | Before 3/17/2020 | | After 3/17/2020 | |
| | | $\beta^L$ | $\beta^U$ | $\beta^L$ | $\beta^U$ |
| S&P500 | Coefficient | -0.358 | 0.747 | -0.468 | 1.357 |
| | t-stat | -44.649*** | 109.904*** | -34.144*** | 102.096*** |
| | | Before 12/01/2020 | | After 12/01/2020 | |
| | | $\beta^L$ | $\beta^U$ | $\beta^L$ | $\beta^U$ |
| S&P500 | Coefficient | -0.399 | 0.801 | 0.22 | 0.745 |
| | t-stat | -55.416*** | 134.914*** | 10.025*** | 33.692*** |
| CSAD | | Before 07/23/2015 | | After 07/23/2015 | |
| | | $\gamma_3$ | | $\gamma_3$ | |

| | | | |
|---|---|---|---|
| S&P500 | Coefficient | 4.547 | -0.634 |
| | t-stat | 2.667** | -33.156*** |
| | | Before 3/17/2020 | After 3/17/2020 |
| | | $\gamma_3$ | $\gamma_3$ |
| S&P500 | Coefficient | -1.524 | -0.3359 |
| | t-stat | -16.328*** | -13.55*** |
| | | Before 12/01/2020 | After 12/01/2020 |
| | | $\gamma_3$ | $\gamma_3$ |
| S&P500 | Coefficient | -1.474 | 0.07 |
| | t-stat | -19.39*** | 3.667*** |

Note: (. p-value<=0.1, * p-value<=0.05, ** p-value<=0.01, ***, p-value<=0.005). Before the market trouble in mid-2015, investors in critical mineral stocks didn't show strong signs of herding, but this behavior became very noticeable afterwards. This herding had already happened before the COVID-19 pandemic hit in March 2020, which might be linked to Indonesia's Nickel Ore Export Ban in January 2020, which continued during the initial crisis. However, a big change happened around December 2020, when good news about vaccines emerged. Instead of herding, investors started to do the opposite (anti-herding), meaning their stocks spread out more during big market moves. How these stocks spread out or bunched together also changed depending on whether the overall market was going up or down across these different periods.

Table 11 suggests that critical mineral equities may exhibit anti-herding behavior during the crisis, as indicated by the higher standard deviation. This increased return dispersion may imply less herding behavior and more independent decision-making among investors. The Russian invasion in February 2022 caused a significant change: before it, investors in critical mineral stocks tended to herd, but after the invasion, they started doing the opposite, with individual stocks moving more independently during big market swings. This suggests the war made investors react differently based on specific risks. For the Mideast crisis in October 2023, it seems herding behavior was present before and continued afterward, though perhaps a bit weaker. Interestingly, after both crises, when the overall market declined, critical mineral stocks tended to spread out more, showing less uniform movement than before the events. The findings provide evidence of an anti-herding dynamic during the crisis, consistent with previous research (Babalos & Stavroyiannis 2015; Stavroyiannis & Babalos 2020). However, detecting anti-herding dynamics can be challenging, particularly without high-frequency trading data. The quality and noise of the data can also affect the accuracy of cross-sectional analysis, and the frequency of the data can affect the reliability of the measure.

Table 11. Results of CSA based on two extreme events (2/22/2022 and 10/7/2023)

| CSSD | | Before 2/22/2022 | | After 2/22/2022 | |
|---|---|---|---|---|---|
| | | $\beta^L$ | $\beta^U$ | $\beta^L$ | $\beta^U$ |
| S&P500 | Coefficient | 0.013 | 0.578 | 0.113 | 0.851 |
| | t-stat | 0.137 | 5.945*** | 3.819*** | 28.853*** |
| | | Before 10/7/2023 | | After 10/7/2023 | |
| | | $\beta^L$ | $\beta^U$ | $\beta^L$ | $\beta^U$ |
| S&P500 | Coefficient | 0.023 | 0.615 | -0.194 | 1.388 |
| | t-stat | 0.459 | 12.017*** | -8.283*** | 65.883*** |
| CSAD | | Before 2/22/2022 | | After 2/22/2022 | |
| | | $\gamma_3$ | | $\gamma_3$ | |
| S&P500 | Coefficient | -0.905 | | 0.166 | |
| | t-stat | -43.432*** | | 7.181*** | |
| | | Before 10/7/2023 | | After 10/7/2023 | |
| | | $\gamma_3$ | | $\gamma_3$ | |
| S&P500 | Coefficient | -0.904 | | -0.076 | |
| | t-stat | -46.978*** | | -3.87*** | |

Note: (. p-value<=0.1, * p-value<=0.05, ** p-value<=0.01, ***, p-value<=0.005). The Russian invasion in February 2022 caused a major change: before it, investors in critical mineral stocks tended to herd, but after the invasion, they started doing the opposite, with individual stocks moving more independently during big market swings. This suggests

that the war led investors to react differently to specific risks. For the Mideast crisis in October 2023, it seems herding behavior was present before and continued afterward, though perhaps a bit weaker. Interestingly, after both crises, when the overall market declined, critical mineral stocks tended to spread out more, showing less uniform movement than before the events.

4.4 Sensitivity analysis

To assess the influence of time horizons on the measurement of herding dynamics, this study conducts a further analysis of data samples covering 2, 4, and 6 months before and after each timeline.

As shown in Table 12, the sensitivity analysis examines changes in herding behavior (as indicated by statistically significant CSAD coefficients) around major market events, demonstrating consistency across 2-, 4-, and 6-month timeframes. The analysis consistently shows increased herding behavior after July 23, 2015, its continuation and strengthening after March 17, 2020 (COVID-19 pandemic), and its emergence following February 22, 2022 (Russia invaded Ukraine). While herding patterns remain detectable, the study shows that their measured strength and the relationship between investment diversity and market declines can vary across timeframes. In some cases, shorter timeframes yielded non-significant results, suggesting that detecting these market dynamics depends on the chosen analysis period.

Table 12. Sensitivity analysis of herding dynamics and leverage effects with 2, 4, and 6 months' time horizons.

| Test | | Time Horizon (months) | | |
|---|---|---|---|---|
| **CSSD ($\beta^L$)** | | 2 | 4 | 6 |
| 07/23/2015 | Before | 0 | 0 | 0 |
| | After | 0 | 0 | 0 |
| 3/17/2020 | Before | 0 | 0 | 0 |
| | After | 0 | 0 | 0 |

| Date | | | | |
|---|---|---|---|---|
| 12/01/2020 | Before | 0 | 0 | 0 |
| | After | -0.03*** | 0.739*** | 0.739*** |
| 2/22/2022 | Before | 0.945*** | 0.986*** | 0.977*** |
| | After | 0.803*** | 0.727*** | 0.732*** |
| 10/07/2023 | Before | 0.502*** | 0.934*** | 0.839*** |
| | After | -0.099*** | -0.199*** | -0.2*** |
| **CSAD ($\gamma_3$)** | | 2 | 4 | 6 |
| 07/23/2015 | Before | -25.338*** | -25.828*** | -27.695*** |
| | After | -95.7*** | 74.224*** | -90.686*** |
| 3/17/2020 | Before | -1.758*** | -2.159*** | -1.624*** |
| | After | -6.834*** | -4.516*** | -3.096*** |
| 12/01/2020 | Before | -1.851*** | -1.729*** | -2.003*** |
| | After | -0.545*** | -1.158*** | -1.035*** |
| 2/22/2022 | Before | 0.021 | 0.157* | -0.076 |
| | After | -0.406*** | -0.45*** | --0.766*** |
| 10/07/2023 | Before | -0.736*** | -0.731*** | -0.035 |
| | After | -0.179*** | -0.055 | -0.058** |

Note: (. p-value<=0.1, * p-value<=0.05, ** p-value<=0.01, ***, p-value<=0.005). The sensitivity analysis examines changes in herding behavior (as indicated by statistically significant CSAD coefficients) around major market events, demonstrating consistency across 2-, 4-, and 6-month timeframes. The analysis consistently shows increased herding behavior after July 23, 2015, its continuation and strengthening after March 17, 2020 (COVID-19 pandemic), and its emergence following February 22, 2022 (Russia invaded Ukraine). While herding patterns remain detectable, the study shows that their measured strength and the relationship between investment diversity and market declines can vary across timeframes. In some cases, shorter timeframes yielded non-significant results, suggesting that detecting these market dynamics depends on the chosen analysis period.

## 5. Discussions

*5.1. Theoretical implications*

This study provides evidence of the advantages of combining CPD and CSA as an event study method in addressing key critiques. For instance, traditional event studies often rely on a single event date, such as the start day of a geopolitical crisis or disaster, which may overlook the latency of market responses. Moreover, event studies focus on a single event in a time series, neglecting other important factors, such as the influence of other markets. Additionally, event studies often assume no information asymmetry and neglect the role of information in shaping market outcomes, leading to biased estimates of event effects and overlooking irrational market behavior and anomalies.

The framework in this study addresses these limitations by offering a more comprehensive approach to investigating investor behavior. Specifically, CPD identifies multiple timelines for each equity in a time series. At the same time, CSA is grounded in the theory about the role of information in shaping market outcomes, which helps to detect irrational market behavior and anomalies. The results of this study address the research questions as follows:

Regarding RQ1, portfolios with higher ESG scores appear to be better insulated from negative market shocks. The analysis points to a clear pattern: equities with lower ESG ratings, such as the REMX fund for rare earth elements, showed greater sensitivity to major market events. This relationship suggests that strong ESG performance may be a factor in building more resilient portfolios in the critical minerals sector, potentially helping to mitigate risk over the long term.

For RQ2, market shocks and extreme events appear to serve as powerful catalysts of investor behavior, triggering noticeable shifts in trading momentum. During broad crises, such as the 2015 oil market shock and the onset of the COVID-19 pandemic, investors tended to exhibit herding behavior, moving together in response to market-wide fear. A different pattern emerged, however, in response to more specific events. The analysis shows a shift toward anti-herding, or more independent decision-making, following positive news like the 2020 vaccine announcements and during the geopolitical turmoil of the 2022 Russian invasion of Ukraine. This distinction indicates that while general economic downturns may foster collective movement, events with more specific risks encourage investors to assess individual assets more carefully.

Finally, for RQ3, the co-movement dynamics among critical mineral portfolios are quite strong, though they fluctuate with market conditions. Investor coordination is most pronounced during market downturns, whereas these relationships show much greater variation during more stable periods. Crucially, the ability to measure these dynamics depends on the chosen time horizon. The sensitivity analysis across 2-, 4-, and 6-month windows shows that while the general patterns of herding remain consistent around key events, the measured strength can vary. In some instances, shorter observation periods did not produce statistically significant results, suggesting that detecting these market behaviors depends heavily on the length of the analytical window.

*5.2. Practical implications*

The results indicate that the timelines identified by CPD are more sensitive to shocks from investments in critical minerals than to those from geopolitical crises. The empirical literature suggests that Russia's invasion of Ukraine led to constrained supply chains and higher prices in global energy markets (Chishti *et al.* 2023), but the data in this study did not exhibit herding behavior on the event date. Instead, the data suggests anti-herding behavior during this crisis. In contrast, the price drops in the oil market and the Chinese stock market crash exhibited herding behavior, as measured by CSA. Anti-herding behavior during geopolitical crises, in contrast to herding in energy markets, can be understood through investor psychology and market co-movement dynamics. During geopolitical crises, such as the Russia–Ukraine conflict, uncertainty and risk perceptions escalate, leading investors to rely more on their private information and analyses rather than following the crowd. This individualistic approach results in anti-herding behavior.

However, energy markets are highly sensitive to geopolitical events due to their direct impact on supply and demand dynamics. During crises, energy market investors display collective behavior, often herding as they anticipate and react to potential supply disruptions. In these markets, geopolitical turmoil can trigger collective behavior among investors, leading to herding as they react to anticipated disruptions in energy supplies. This herding behavior in energy commodity futures markets has been documented during periods of turmoil. Thus, while geopolitical crises may prompt individualistic behavior in broader stock markets due to diverse interpretations of risk, the direct and immediate implications of such events on energy markets foster a collective response among investors.

These findings provide valuable insights for companies, investors, and regulators regarding systematic risk and informed investment strategies. This paper pioneers a comprehensive approach to exploring the interconnections between critical minerals investment and other market shocks, shedding light on the dynamic risks they pose.

The analysis shows that equities with low ESG scores, such as REMX, are more sensitive to market shocks after the five timelines. This evidence underscores the importance of ESG disclosure and the development of sustainable supply chains for critical mineral exploration and production.

The identified structural changes before and after the five timelines highlight the need for heightened awareness among investors and regulators of the contagion effects of shocks from energy markets and the uncertain risks from other markets. Spillovers from other markets can impact price fluctuations and short-term volatility in critical minerals portfolios.

Collecting quantitative data from diverse markets within the same time scale or period is recommended to enhance future research. Additionally, evolving international regulatory policies may introduce further complexity to systemic shocks, potentially eluding capture by existing models or methods. Therefore, future research should explore new models or measurements that account for these evolving risks in data collection and model design.

*5.3. Limitations and future research directions*

While the proposed analytics framework offers several advantages in assessing herding dynamics, it has limitations. The framework's effectiveness depends heavily on data quality, underscoring the need for high-quality data to ensure accurate results. To address the limitations of the CPD and CSA models, future research should explore the impact of external shocks, such as social media exposure, government regulations, and policies, on herding dynamics. This could provide a more comprehensive understanding of the complex interactions between market participants and external factors.

## 6. Conclusion

This study highlights the advantages of change-point detection over event-day analysis, as it provides a more precise timeline for investigating investor behavior and offers greater explanatory power. This study provides stronger evidence of herding behavior by capturing the latency of market responses to significant events. This study identifies significant herding dynamics in the investment in critical mineral equities throughout the entire period, particularly around market-moving events in the energy market. Significant herding behavior after the 2015 price drop in the world oil market suggests a phenomenon driven by investors in renewable energy development. This study shows that herding dynamics intensify during oil market declines, driven by specific investor traits. By incorporating CSSD and CSAD metrics, this research quantifies the consensus or dispersion among market participants in the energy market. Empirical evidence suggests anti-herding behavior among critical mineral equities during the geopolitical crisis, but prompts herding behavior on the timeline associated with the energy market.

The findings of this study provide actionable insights for market participants and regulators, offering multiple benefits. Firstly, the study's insights into herding dynamics can significantly enhance the performance of Modern Portfolio Theory (MPT) for portfolio optimization and asset allocation in renewable energy development. By incorporating herding dynamics into MPT, investors can create more robust portfolios that adapt to changing market conditions. Secondly, the leverage effect analysis conducted in this study provides valuable insights for investors to assess MPT's sensitivity to asymmetric shocks. This analysis enables investors to identify stocks with high non-diversifiable risks, such as geopolitical crises, making MPT more effective in managing portfolio risk over time. Investors can develop more effective strategies to mitigate investment losses and optimize their portfolios by identifying critical mineral equities most vulnerable to large drawdowns during asymmetric shocks.

**Disclosure of interest**: The Author has no conflict of interest

**Statement of Funding:** No funding was received

**Data Availability Statement**: The data containing ETFs' daily closing prices are available at 10.6084/m9.figshare. 28510169.

**List of Abbreviations and Nomenclature**

**Abbreviations**

**Abbreviation** **Full Name/Description**

| Abbreviation | Full Name/Description |
|---|---|
| ADF | Augmented Dickey-Fuller (test for stationarity) |
| ARCH | Autoregressive Conditional Heteroskedasticity |
| CPD | Change Point Detection |
| CRSP | Center for Research in Security Prices |
| CSA | Cross-Sectional Analysis |
| CSAD | Cross-Sectional Absolute Deviation |
| CSSD | Cross-Sectional Standard Deviation |
| EPI | Environmental Performance Index |
| ESG | Environmental, Social, and Governance |
| ETF | Exchange-Traded Fund |
| GEI | Government Effectiveness Index |
| GFC | Global Financial Crisis |
| GJR-GARCH | Glosten-Jagannathan-Runkle Generalized Autoregressive Conditional Heteroskedasticity |
| IEA | International Energy Agency |
| LM | Lagrange Multiplier (statistic) |
| MI | Mutual Information |
| MPT | Modern Portfolio Theory |
| MSCI | Morgan Stanley Capital International |
| PELT | Pruned Exact Linear Time (algorithm) |
| RQ | Research Question |
| SPI | Social Progress Index |
| SPX | S&P 500 Index (Ticker Symbol) |
| WHO | World Health Organization |
| WTI | West Texas Intermediate (crude oil benchmark) |

**Appendix**

### A. Pruned Exact Linear Time (PELT) algorithm

The Pruned Exact Linear Time (PELT) algorithm is a dynamic programming method for efficiently detecting multiple changepoints in time series data. It can identify points where the statistical properties of the data shift, optimizing both accuracy and computational efficiency (Killick *et al.* 2012; Wambui *et al.* 2015).

PELT seeks to minimize a cost function that balances the model's fit to the data with a penalty term to prevent overfitting by adding unnecessary changepoints. The objective is to find the set of changepoints, denoted as $\{\tau_1, \tau_2, \cdots, \tau_m\}$, $\tau_1 = 0$, $\tau_{m+1} = n$, which minimizes the following penalized cost function (Killick *et al.* 2012):

$$Min \sum_{i=1}^{m+1}\left[C(y_{(\tau_{i-1}+1):\tau_i})\right] + \beta f(m) \tag{1}$$

, where $y_{1:n} = (y_1, \cdots, y_n)$ represents the ordered sequence of observed data, m is the number of change points, $C(y_{(\tau_{i-1}+1):\tau_i})$ denotes the cost function for measuring the fit of the segment, and $\beta f(m)$ denotes a penalty term that controls the trade-off between the number of changepoints and the model's fit.

The PELT algorithm improves computational efficiency through a pruning mechanism. At each time point *t*, it evaluates potential changepoints and eliminates those that cannot be part of the optimal solution based on a pruning condition. This condition ensures that only promising candidates are considered, significantly reducing the computational burden.

**PELT algorithm**

*Initialization: a time series dataset, d, Set $C(0) = -\beta$ and initialize an empty set of changepoints.*
**Step 1:** For each time point t from 1 to n:
    1: Compute the cost for ending a segment at t.
    2: Apply the pruning condition to discard unlikely changepoint candidates.
    3: Update $C(t)$ with the minimum cost found.
    4: Store the optimal previous changepoint for Step 2.
**Step 2**: After processing all time points, backtrack from the end to identify the optimal set of changepoints.

PELT can achieve a linear computational cost with respect to the number of data points, making it highly efficient for large datasets.

### B. Mutual information function

Mutual information (MI) quantifies the amount of information obtained about one random variable through the observation of another, effectively measuring the dependence between two variables (Learned-Miller 2013).

For discrete random variables $X$ and $Y$ with a joint probability mass function $p(x, y)$ and marginal probability mass functions $p(x)$ and $p(y)$, respectively, the mutual information is defined as:

$$MI\ (X;Y) = \sum_{x \in X} \sum_{y \in Y} p(x,y) \log \left( \frac{p(x,y)}{p(x)p(y)} \right) \qquad (2)$$

Mutual information is inherently non-negative and equals zero if and only if the variables are statistically independent, indicating no mutual dependence. It is symmetrical, meaning $MI\ (X;Y) = MI\ (Y;X)$. These relationships highlight that mutual information captures the reduction in uncertainty of one variable given knowledge of the other.

### C. CSSD and CSAD metrics of cross-sectional analysis

The CSSD metric, introduced by Christie and Huang (1995), is a valuable statistical tool for analyzing herding behavior in extreme market conditions, particularly in financial markets. Herding behavior refers to the tendency of individuals to follow the majority or consensus rather than make independent decisions. The CSSD metric is useful for quantifying the dispersion of individual behavior or opinions within a group, specifically in the context of critical minerals equities. However, the CSAD metric is better suited to detecting herding behavior during less extreme market movements or asymmetric herding dynamics.

The CSAD metric, widely used in finance, is a robust measurement tool for estimating firm-level volatility of individual assets within a cross-sectional framework (Chang et al. 2000). In the context of critical mineral equities, the CSAD metric is particularly valuable due to the unique roles and functions of technology development in the renewable energy sector. By analyzing herding behavior using the CSAD metric, researchers can gain insights into the dynamics of critical mineral equities and the impact of technology development on market behavior.

Herding dynamics can lead to volatility clusters, in which periods of high volatility are followed by periods of low volatility, producing a clustering effect. This phenomenon is characterized by returns that exhibit oscillations in magnitude, alternating between large and small values at a low frequency over an extended period. To capture the persistent nature of volatility fluctuations associated with herding dynamics in critical mineral equities, this study utilizes a GJR-GARCH model (Glosten et al. 1993). The CSSD metric, as introduced by Christie and Huang (1995), is employed to quantify the dispersion of individual behavior or opinions within a group and is defined as follows:

$$CSSD_t^2 = \frac{\sum_{i=1}^{N}(R_{i,t}-R_{m,t})^2}{N-1} \qquad (3)$$

The CSSD metric, denoted as $CSSD_t$, measures the average distance between individual asset returns and the overall market return. Specifically, it calculates the average proximity of individual returns to the realized average, where $R_{m,t}$ represents the average return of assets in the aggregate market portfolio at time *t*, and $R_{i,t}$ signifies the return of an individual asset *i* at time *t*. The number of assets in cross-sectional observations is denoted as *N*. In rational asset pricing models, $CSSD_t$ increases with the absolute difference between individual asset returns and the market return, $|R_{i,t} - R_{m,t}|$, reflecting disparities among individual assets in their responsiveness to market movements. However, in the presence of herding dynamics, individual assets tend to move in tandem, resulting in smaller deviations from the overall market return. This phenomenon challenges the assumptions of rational asset pricing models. To investigate the impact of herd behavior during periods of significant market movements, Christie and Huang (1995) developed a regression model to estimate $CSSD_t$, using two binary variables, $D_t^L$ and $D_t^U$. These variables are designed to capture differences in investor behavior during exceptional market conditions, such as the distribution's highly skewed lower and upper extremities. The regression model for $CSSD_t$ is formulated as follows:

$$CSSD_t = \alpha + \beta^L D_t^L + \beta^U D_t^U + \varepsilon_t \tag{4}$$

The presence of herding behavior can be inferred from the regression model, where a negative and statistically significant coefficient $\beta^L$ and $\beta^U$ would indicate that investors are following the crowd. A low CSSD value suggests a strong consensus among investors, indicating that they are making similar decisions or taking similar actions. Conversely, a high CSSD value suggests investors make more diverse decisions, indicating a lack of herding behavior. Changes in CSSD values over time can indicate shifts in investor behavior, such as a transition from herding to more independent decision-making or vice versa. Additionally, analyzing CSSD values surrounding events such as earnings releases or economic reports can provide insights into the impact of information cascades on investor behavior, offering empirical evidence to support the Information Cascades Theory.

The CSAD metric provides a comprehensive measure of asset's volatility by calculating the average absolute deviation of its returns from the cross-sectional mean. This singular metric effectively captures the overall dispersion of individual behavior around the mean, serving as an essential indicator of consensus or divergence among market participants. Therefore, CSAD values provide empirical evidence that supports the Reputational Herding Theory, which suggests that fund managers or analysts may replicate the investment decisions of others to protect their reputations, even if they possess different private information about individual assets.

The CSAD metric, employed in this study to analyze herding behavior in critical mineral equities, is defined by Chang et al. (2000) as follows:

$$CSAD_t = \alpha + \gamma_1 R_{m,t} + \gamma_2 R_{m,t}^2 \tag{5}$$

The CSAD metric is calculated as a function of the return on the aggregate market portfolio, $R_{m,t}$, on day $t$. Notably, the quadratic relationship between $CSAD_t$ and $R_{m,t}$ suggests that $CSAD_t$ reaches its maximum value when $R_{m,t} = -(\gamma_1/2\gamma_2)$. . This implies that a lower $CSAD_t$ value indicates a higher level of consensus or herding behavior among critical mineral equities, suggesting that investors are closely aligned in their actions. Conversely, a higher $CSAD_t$ value suggests more significant divergence, indicating a lack of consensus among investors.

Given the potential for asymmetry in herding dynamics across different market conditions, the CSAD metrics for both upward and downward markets are presented below, as outlined in Chang et al. (2000):

$$CSAD_t^U = \alpha + \gamma_1^U |R_{m,t}^U| + \gamma_2^U (R_{m,t}^U)^2 + \varepsilon_t \tag{6}$$

$$CSAD_t^D = \alpha + \gamma_1^D |R_{m,t}^D| + \gamma_2^D (R_{m,t}^D)^2 + \varepsilon_t \tag{7}$$

To capture the distinct dynamics of both upward and downward market conditions, this study incorporates an additional term, $\gamma_2 |R_{m,t}|$, into Equation (3), thereby enhancing the model's ability to account for market asymmetry

$$CSAD_t = \alpha + \gamma_1 R_{m,t} + \gamma_2 |R_{m,t}| + \gamma_3 R_{m,t}^2 \tag{8}$$

A statistically significant and negative $\gamma_3$ coefficient would indicate the presence of herding behavior, suggesting that investors tend to follow the crowd rather than make independent decisions.